\documentclass[12pt, a4paper]{article}

\usepackage{amsmath}
\usepackage{amsfonts}
\usepackage{amssymb}
\usepackage{graphicx, rotating}
\usepackage{epstopdf}
\usepackage{epsfig}
\usepackage{latexsym}
\usepackage{graphicx}
\usepackage{color}
\usepackage{amsmath,amssymb}
\usepackage{cite}
\usepackage{slashed}
\usepackage{hyperref}
\hypersetup{colorlinks, citecolor=bluscuro, linkcolor=black, urlcolor=bluscuro}
\definecolor{rossos}{cmyk}{0,1,1,0.55}
\definecolor{bluscuro}{rgb}{0.15, 0.2, .85}
\definecolor{bluchiaro}{cmyk}{1,.3,0.,0.1}

\newcommand{\eq}[1]{Eq.~(\ref{#1})}

\newcommand{\deslash}{\!\not\! \partial}
\newcommand{\be}{\begin{equation}}
\newcommand{\ee}{\end{equation}}
\newcommand{\bea}{\begin{eqnarray}}
\newcommand{\eea}{\end{eqnarray}}

\newcommand{\TeV}{\,\mathrm{TeV}}
\newcommand{\GeV}{\,\mathrm{GeV}}

\begin{document}

%FRONTPAGE2%%%%%%
\begin{titlepage}
\begin{flushright}
%UAB-FT--\\
November 2012
\end{flushright}
\vspace{.3in}

\begin{center}
\vspace{1cm}

{\Large \bf

Is the 125 GeV Higgs the superpartner of a neutrino?\\

}

\vspace{1.2cm}

{\large Francesco Riva\,$^{a,b}$, Carla Biggio\,$^{b,c}$ and Alex Pomarol\,$^d$\\}
\vspace{.8cm}
{\it {$^a$\,  Institut de Th\'eorie des Ph\'enom\`enes Physiques, EPFL,1015 Lausanne, Switzerland}\\
{\it {$^b$\, IFAE, Universitat Aut{\`o}noma de Barcelona,
   08193~Bellaterra,~Barcelona}}\\
{\it {$^c$\, Dipartimento di Fisica, Universit\`a di Genova, 16146
    Genova, Italy}}\\
    {\it {$^d$\, Dept.  de F\'isica, Universitat Aut{\`o}noma de Barcelona, 08193~Bellaterra,~Barcelona}}\\

}

\vspace{.4cm}

\end{center}
\vspace{.8cm}

\begin{abstract}
\medskip
\noindent
\end{abstract}
Recent LHC searches have provided   strong   evidence for the Higgs,  a boson whose gauge quantum numbers  coincide with  those of a SM fermion, the neutrino.
This raises  the  mandatory  question of whether Higgs and neutrino can be related  by  supersymmetry.
We study this possibility  in a model in which  an approximate  $R$-symmetry    acts  as a  lepton number.
We show that   Higgs  physics resembles  that of the  SM-Higgs with the exception of  a novel
 invisible decay into Goldstino and neutrino with a branching fraction that can be as large as   $\sim 10\%$. 
Based on  naturalness criteria, 
 only stops  and sbottoms are required  to be lighter than the TeV
with  a phenomenology   dictated by the $R$-symmetry. They
have  novel decays 
 into quarks+leptons that could be 
 seen at the LHC,    allowing    to  distinguish these scenarios  from the ordinary MSSM.  
\bigskip

\end{titlepage}

%%%%%%%%%%%%%%%%%%%%%%%%%%%%%%%%%%%%%%%%
%%%%%%%%%%%%%%%%%%%%%%%%%%%%%%%%%%%%%%%%
%%%%%%%%%%%%%%%%%%%%%%%%%%%%%%%%%%%%%%%%
%%%%%%%%%%%%%%%%%%%%%%%%%%%%%%%%%%%%%%%%
\section{Introduction}

The LHC has recently  reported strong  experimental evidence for  the existence of the  Higgs particle.
This is  the  first discovered boson  whose   gauge quantum numbers  are the same as  those of an existing fermion, the neutrino.  
It is therefore tempting to speculate on the most minimal realization
of  supersymmetry, needed to  protect the Higgs mass,
   corresponding to a situation where the Higgs and a  neutrino (any of the three)  belong to the same supermultiplet.
In this article we study requirements and implications of this possibility.

If the  Higgs is the neutrino  superpartner,
an  approximate  $R$-symmetry $U(1)_R$,  that  acts as a lepton symmetry~\footnote{The idea of an $R$-symmetry as a lepton symmetry was first proposed in 
\cite{Fayet:1976et}.  In this  original realization, however,  the particle spectrum was not realistic.},  is necessary \cite{Gherghetta:2003he,Frugiuele:2011mh}.
This is needed to  provide the neutrino with  an approximate conserved  lepton number  that protects its mass,
while leaving its supersymmetric partner, the Higgs, without lepton charge. In this way, the latter
can  acquire a nonzero vacuum expectation value (VEV) and break all  symmetries under which it is charged, without breaking lepton number.
As we will show, there are important implications of this  approximate $R$-symmetry.
Since   the gravitino   is  $R$-charged,
 the Higgs can  decay into a neutrino and a gravitino,  with a branching ratio
 that can be as large as $10\%$.
This gives  an invisible decay width to the Higgs  
that could  be indirectly  detected  by  measuring   a  small reduction of all its  visible branching ratios.
 Gauginos must  get Dirac, rather than Majorana, masses  and 
 the wino mass must lie above the TeV in order to  avoid large corrections to charged leptons couplings \cite{Frugiuele:2011mh}.
Therefore  gauginos are  not expected  to be detectable during the
first years of  the LHC running. 
Another requirement of the model is that,
if no extra  Higgs superfields are present (and hence no Higgsinos),    the up-quark Yukawa couplings
 must arise  from  a supersymmetry-breaking term.
Interestingly, however, we will show
that the soft-mass of the Higgs is insensitive (at the one-loop level)
to this 
supersymmetry-breaking term
  that  can have its origin in physics above the TeV,
as  we propose  in the Appendix.

In a  bottom-up approach to supersymmetry based on naturalness criteria,
models with the Higgs  as a neutrino superpartner  have
 the most minimal low-energy   supersymmetric spectrum,
  since no  Higgsinos are present (hence avoiding  the  infamous  $\mu$ problem).
Below the TeV, only stops and sbottoms are  required, but
with a  phenomenology very different   from that of the Minimal
Supersymmetric Standard Model (MSSM).
In particular, stops  and sbottoms  exhibit leptoquark decays: 
$\tilde t_L\rightarrow b\bar l^-,t\bar \nu$, 
$\tilde t_R\rightarrow t\nu$, while $\tilde b_L\rightarrow b\bar \nu$.
These decay channels can compete with decays into gravitino (a channel
that is also present in the MSSM with low-scale supersymmetry breaking), thus allowing to differentiate this model from the MSSM.
We  will discuss the precise branching ratios, the present bounds and future searches to 
 discriminate between these scenarios. 
  If light enough to be produced at the LHC, we will show that first and second generation squarks  could decay dominantly into 3-bodies including quarks, leptons and gauge/Higgs bosons, providing then distinctive novel signatures to be searched at the LHC.

%%%%%%%%%%%%%%%%%%%%%%%%%%%%%%%%%%%%%%%%
%%%%%%%%%%%%%%%%%%%%%%%%%%%%%%%%%%%%%%%%
%%%%%%%%%%%%%%%%%%%%%%%%%%%%%%%%%%%%%%%%
%%%%%%%%%%%%%%%%%%%%%%%%%%%%%%%%%%%%%%%%
\section{The Higgs as a lepton superpartner}

We consider a model that, differently from the MSSM, does not
contain the two Higgs superfields $H_{u}$ and $H_{d}$.  Instead, the
SM scalar Higgs doublet is assumed to be one of the three lepton
superpartners.  The corresponding chiral superfield is denoted by
\be H\equiv L_3=(H,l_L)\, , \ee where we label by $l_L=(l^-_L,\nu_L)$
one of the three left-handed leptons, either the electron, muon or tau doublet.  The other two
 are embedded  in the chiral superfields
$L_{1,2}\equiv(\tilde L_{1,2},l_{L_{1,2}})$.  The full spectrum of the
theory is given in table~\ref{table1}.  Notice that this theory does
not have Higgsinos and is of course anomaly free, since the only extra
fermions beyond the SM are all in adjoint representations. 

\begin{table}[top]
\centering 
\begin{tabular}{|c|c|c|}
\hline
 & 
\textbf{$  SU(3)_c\times SU(2)_L\times U(1)_Y $}& 
\textbf{$U(1)_R$} \\ 
\hline 
$Q$ & \  \ \ \ \ \ \ $ (3,2)_{\frac{1}{6}}$ \ \ & $1+B $\\ 
$U$ & \  \ \ \ \ \ \ $ (\bar 3,1)_{-\frac{2}{3}}$ \ \ & $1-B$ \\  
$D$ & \  \ \ \ \ \ \ $ (\bar 3,1)_{\frac{1}{3}}$ \ \ & $1-B$ \\ 
$L_{1,2}$ & \  \ \ \ \ \ \ $ (1,2)_{-\frac{1}{2}}$ \ \ & $1-L$ \\ 
$E_{1,2}$ & \  \ \ \ \ \ \ $ (1,1)_{1}$ \ \ & $1+L $\\ 
$H\equiv L_3$ & \  \ \ \ \ \ \ $ (1,2)_{-\frac{1}{2}}$ \ \ & 0 \\
$E_3$ & \  \ \ \ \ \ \ $ (1,1)_{1}$ \ \ & 2 \\  
$W^\alpha_a$ & \  \ \ \ \ \ \ $ (8,1)_0+(1,3)_0+(1,1)_0 $\ \ & 1 \\ 
\hline
$ \Phi_a $ &  \  \ \ \ \ \ \ $ (8,1)_0+(1,3)_0+(1,1)_0$ & 0 \\ 
\hline
$ X\equiv \theta^2F $ &  \  \ \ \ \ \ \ $  (1,1)_0$ & 2\\ 
\hline\end{tabular} 
\caption{\emph{Superfield  content  and charge assignments under the SM gauge group and the $U(1)_R$ symmetry.
 The value of the $R$-charge ($q_R$) corresponds to the charge of the superfield and the scalar component, while the fermion component has charge $q_R-1$ and the F-term has charge $q_R-2$. 
$B$ and $L$  are arbitrary charge assignments.
 }}
\label{table1}
\end{table} 

Any theory beyond the SM  must preserve an approximate  lepton number  in order to avoid
large neutrino masses. In our model this lepton symmetry cannot commute with supersymmetry,
otherwise  the Higgs $H$, being in the same supermultiplet as the leptons, would carry lepton number and this
  would be broken when the Higgs gets a VEV.
For this reason lepton number can only be defined 
as an $R$-symmetry $U(1)_R$ under which $H$ is neutral but $l_L$ is charged.
The $R$-charges  for this model are given in table~\ref{table1}. 
Few comments are in order.
Since
gauginos must carry nonzero $R$-charges, they cannot 
get  Majorana masses.  Nevertheless, they can get Dirac-type masses
by  marrying  with additional fermions coming from adjoint chiral superfields $\Phi_a$.
Notice also that there is a certain freedom in the symmetry properties of  quarks and
$l_{ 1,2}$ leptons, depending on whether or not they transform under the $U(1)_R$ ($B,L\not= 0$). 
A non-vanishing charge $B\neq 0$ corresponds to a non-vanishing $U(1)_R$ charge for protons and neutrons
 that can be used to protect  proton decay.   Indeed, 
for $B\neq  |L|$ the proton decays to   neutrinos or positrons are  forbidden by the $R$-symmetry,
as well as the decay  into  (anti)gravitinos (of $R$-charge $\mp 1$)
  if  $|B|\neq 1/3$.
Also for $L  \not=0$  the $R$-symmetry  protect the   masses    of all the three neutrinos,
and for $L  \not=1$   the  superpotential terms   $L_iL_jE_k$ and
$Q_iL_jD_k$, which are strongly constrained by lepton-flavor violating  processes \cite{Frugiuele:2012pe,Barbier:2004ez}, are not allowed.

Working  with   $B\neq  1/3$ and $L\neq 1$,  the only superpotential terms that can be written  in this model at the renormalizable level are, including only matter fields, 
\begin{equation}\label{W}
W= Y_d\, HQ  D+Y_{e\, ij}\, HL_i E_j\, ,
\end{equation}
where indexes $i,j=1,2$ are summed over and $Y_d$ is a matrix in flavor space. 
As it stands, the superpotential \eq{W} does not generate up-type quark masses, gaugino masses, nor a mass for the $l^-_L$ lepton (the latter is forbidden since $SU(2)_L$ indices in \eq{W} are summed antisymmetrically, meaning that the term $HHE_3$ vanishes). These must originate as supersymmetry-breaking effects. 
We can write these in a supersymmetry preserving notation by means of a spurion field $X$, whose $F$-component is nonzero
$X=\theta^2 F$. To preserve the $R$-symmetry, $X$ must have $R$-charge $2$.
 The masses of the up-type quarks can be written as
\begin{equation}\label{topmass}
\int d^4\theta\ y_u\frac{  X^\dagger}{M} \frac{  H^\dagger Q U}{\Lambda}  = \int d^2\theta\ Y_uH^\dagger Q U\, ,
\end{equation}
where
$y_u$ are dimensionless couplings and $Y_u\equiv y_uF/(M\Lambda)$ are the Yukawa couplings of the up-type quarks.
Notice that we have defined   two scales, $M$ and $\Lambda$, that could have different origin:
 $M$ is the scale at which the supersymmetry-breaking effects are mediated  to the SM superpartners, while
 $\Lambda$  is the scale at which the higher-dimensional operator \eq{topmass} 
is generated. Explicit examples for the origin of this operator are given in the Appendix.
Since,  as we will   see, the soft masses of  SM superpartners  are of order $F/M$, 
 naturalness requires   $F/M\lesssim$ TeV.
On the other hand,
since the Yukawa coupling of the top is of order one, $Y_t\sim 1$,
we need  $\Lambda\sim y_u F/M\lesssim 4\pi\,   {\rm TeV}$.
The mass for the lepton $l_L^-$ can originate from supersymmetry-breaking terms as well. 
Indeed, we can  have\cite{Grant:1999dr}
\begin{equation}\label{Leptonamass}
 \int d^4\theta\ y_3\frac{X^\dagger  X}{M^2} \frac{H D^\alpha H D_\alpha E_3}{\Lambda^2} 
 \, ,
\end{equation}
where $D_\alpha$ is
 the superspace derivative.
This  term generates  a  Yukawa    for 
$l_L$    equal to
$Y_l= y_3F^2/(M^2\Lambda^2)$.

Gauge boson superpartners must also get masses from supersymmetry-breaking terms.
Dirac-type gaugino masses can arise from
\begin{equation}
\label{eq:mgauginos}
\int d^2\theta\ \frac{D^\alpha X}{M}W^{a}_\alpha \Phi_{a}\, ,
\end{equation}
that induces gaugino masses of order $F/M$.
There are important  constraints on these masses
since, after electroweak symmetry breaking (EWSB),  charged winos mix with $l^-_L$  \cite{Frugiuele:2011mh} as they have
equal $R$-charges. This mixing 
affects the coupling of  $Z$ to $l^-_L$ as
\begin{equation}
\delta g^l_{V,A}=- \frac{m_W^2}{M_{\tilde{W}}^2+2 m_W^2}\, ,
\end{equation}
where $M_{\tilde{W}}$ is the wino mass.
Taking  the bounds on $\delta g^l_{V,A}$ from \cite{ParticleData:1970aa}, we obtain  
at 99\%C.L.  the following  lower bounds~\footnote{Charged current universality is also affected  \cite{Pich:1997hj} but this puts only a mild constraint on the bino mass $M_{\tilde{B}}\gtrsim 500 \GeV$.}
\be\label{massgauginobound}
M_{\tilde{W}}\gtrsim\left\{\begin{array}{cl}2.5\TeV& l^-_L= e_L \\2\TeV & l^-_L=\mu_L \\  1.8\TeV & l^-_L=\tau_L\end{array}\right. ,
\ee
which can be satisfied for  $F/M\sim$ TeV.
The term \eq{eq:mgauginos} does not give mass to the imaginary part of the scalar component  in $\Phi_a$,
but  this can arise from  other supersymmetry-breaking terms such as
$\int d^4\theta\  XX^\dagger \Phi_{a}^2/M^2$.

Finally, the $R$-symmetry forbids the appearance of  supersymmetry-breaking  trilinear $A$-terms,
implying that the stop one-loop corrections   to the Higgs mass are not enough to
give $m_h\sim 125$ GeV for  stop masses  below the TeV, as required by naturalness.
New contributions to the $D$-term Higgs quartic are  then needed.
These  can come from supersymmetry-breaking interactions  of the type
\begin{equation}
\int d^4\theta\  \lambda_H \frac{X^\dagger X}{M^2}\frac{|H|^4}{\Lambda^2}= \delta \lambda_h\, h^4+\dots  \, ,
\label{extraquartic}
\end{equation}
that could either be induced from integrating heavy vector fields (of mass $\Lambda$) that would give extra $D$-terms,
or from coupling $H$ directly to the supersymmetry-breaking mediators 
\cite{Azatov:2011ht}.
In order to obtain
$m_h=125\GeV$, we need  $\delta \lambda_h\sim 0.015$. 

The $R$-symmetry cannot be an exact symmetry of the model.
  In order to adjust the cosmological constant  to (almost)  zero,
 a gravitino Majorana mass of order
\be
m_{3/2}\sim \frac{F}{M_P}\simeq 10^{-4}\ {\rm eV} \left( \frac{\sqrt{F}}{\rm 2\ TeV}\right)^2\, ,
\label{gravitino}
\ee
is needed. This breaks the $R$-symmetry explicitly and generates neutrino masses of order $m_{3/2}$,
which can be in agreement with the experimental limits for $m_{3/2}\lesssim 10$ MeV
(or, equivalently, $\sqrt{F}\lesssim 10^{7}$ GeV)
\cite{Barbier:2004ez,Frugiuele:2011mh,Bertuzzo:2012su}.
These upper-bound however can be evaded 
in theories with emergent global supersymmetry   \cite{Luty:2002ff,Gherghetta:2003he,Sundrum:2009gv}
in which 
the supersymmetric SM (or part of it)  arises from a strong sector at high-energies.
The $R$-symmetry  is  an accidental symmetry of these models   
 not broken at order $m_{3/2}$ but by much smaller effects.
 The gravitino mass   can then be much  heavier  than TeV,  and then irrelevant for  the phenomenology of the model.
Having this in mind, we will consider scenarios
in which either a neutrino or the gravitino  is  the lightest $R$-charged  particle.

Summarizing, the Higgs as a lepton superpartner requires, at least, the supersymmetry-breaking operators
 Eqs.~(\ref{topmass}), (\ref{Leptonamass}), (\ref{eq:mgauginos}) and  (\ref{extraquartic}).
We will not elaborate here on how these supersymmetry-breaking terms could arise from  a specific renormalizable theory, but just postulate that 
this is the case and study their implications.
Nevertheless, we give in the Appendix  possible ultraviolet (UV) completions of   these Higgsinoless models.

%%%%%%%%%%%%%%%%%%%%%%%%%%%%%%%%%%%%%%%%
%%%%%%%%%%%%%%%%%%%%%%%%%%%%%%%%%%%%%%%%
\subsection{The most natural supersymmetric spectrum}\label{sec:Naturaleness}

\begin{figure}[top]
\begin{center}
\includegraphics[width=5cm]{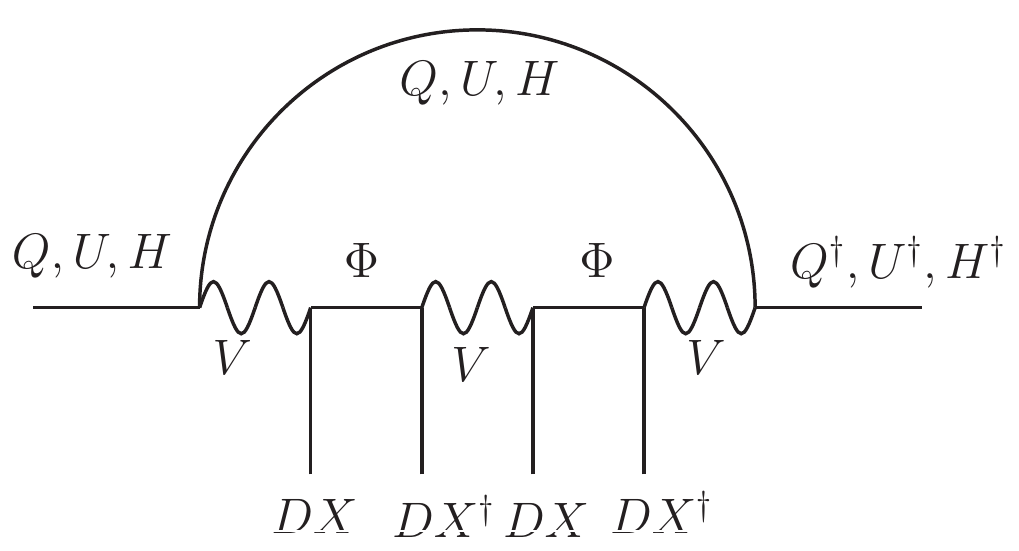}
\end{center}
     \caption{\footnotesize \emph{Gaugino loop contribution to scalar soft masses  arising  from \eq{eq:mgauginos}.}}\label{fig:feyndiagDivergence0}
\end{figure}

\begin{figure}[t]
\begin{center}
  \includegraphics[width=5cm]{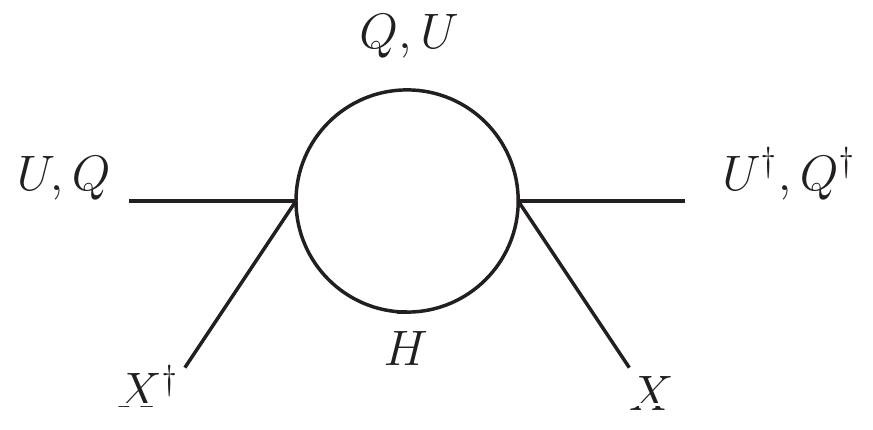}
\end{center}
     \caption{\footnotesize \emph{Feynman supergraphs  
   arising from \eq{topmass}   contributing to the  squark soft masses.}}
     \label{fig:feyndiagDivergence}
\end{figure}

\begin{figure}[t]
\begin{center}
\includegraphics[width=4cm]{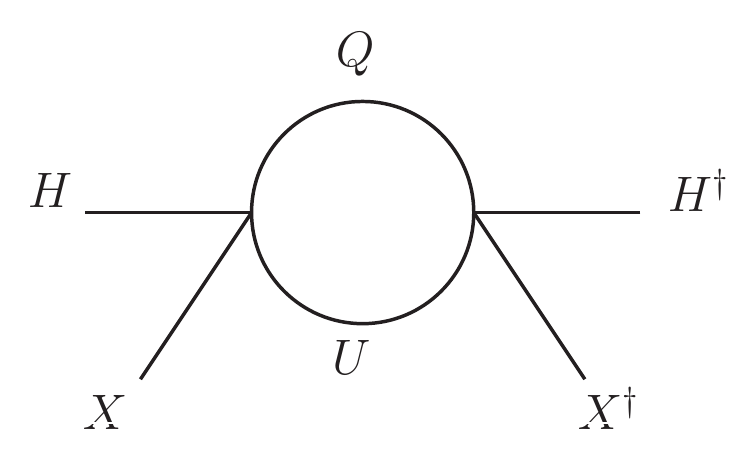}\hspace{1cm}
    \includegraphics[width=4cm]{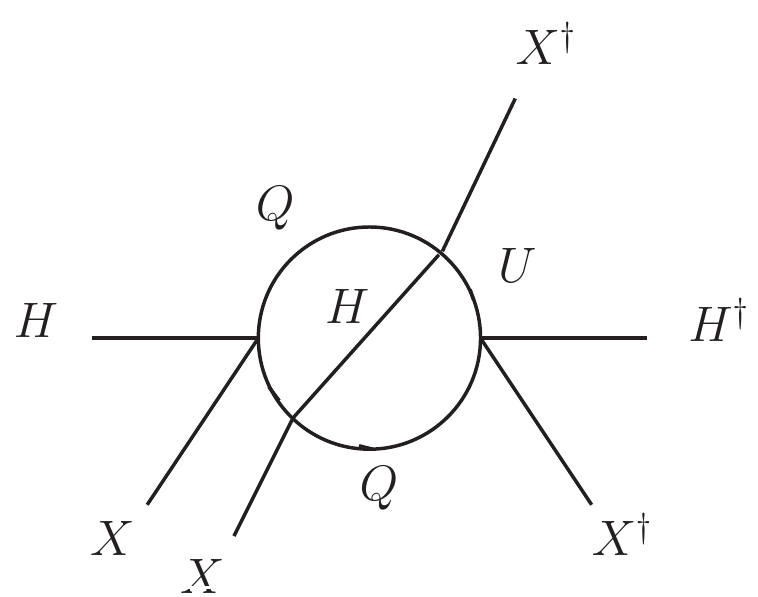}\hspace{1cm}
  \includegraphics[width=4cm]{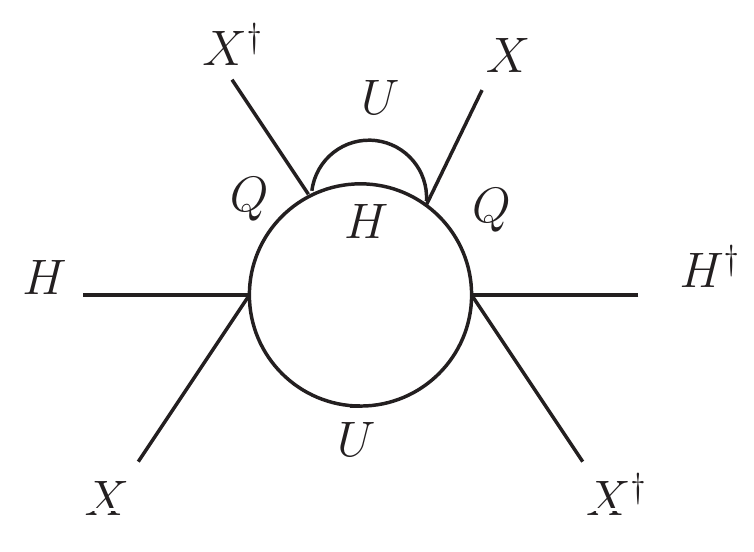}
\end{center}
     \caption{\footnotesize \emph{Feynman supergraphs arising from \eq{topmass} 
     potentially contributing to the Higgs soft-mass.}}\label{fig:feyndiagDivergence2}
\end{figure}

\begin{figure}[top]
\begin{center}
\includegraphics[width=3cm]{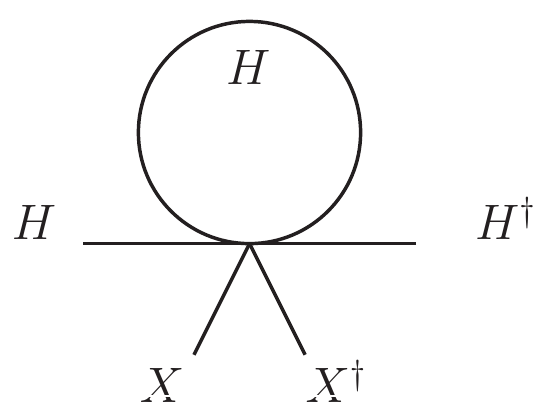}
\end{center}
     \caption{\footnotesize \emph{Feynman supergraphs  arising from \eq{extraquartic}  contributing to the Higgs soft-mass.}}\label{fig:feyndiagDivergence3}
\end{figure}

The presence of the operators Eqs.~(\ref{topmass}), (\ref{Leptonamass}), (\ref{eq:mgauginos})  and  (\ref{extraquartic}), 
 generates at the loop level
other  operators. 
Therefore  it is   natural  in a quantum field theory  to include all of them.
For example, from loop effects,  as depicted in figs.~\ref{fig:feyndiagDivergence0}-\ref{fig:feyndiagDivergence3}, 
the following terms are expected:
\begin{eqnarray}
\label{eq:mstop}
 \int d^4\theta\ \left\{
 g_{Q}\frac{X^\dagger X}{M^2} Q^\dagger Q 
+ g_{U}\frac{X^\dagger X}{M^2} U^\dagger U+
g_{H}\frac{X^\dagger X}{M^2} H^\dagger H\right\}\ ,
\end{eqnarray}
and similarly for the leptons $L_{1,2}$.
These terms   give supersymmetry-breaking (soft) masses for the Higgs $m^2_H=g_HF^2/M^2$ and squarks $m_{Q,U}^2=g_{Q,U}F^2/M^2$.
It is then crucial  to estimate their size, in order to identify the most natural superpartner mass-spectrum
 of the model.
Let us start with the gauge contribution arising from the supersymmetry-breaking term \eq{eq:mgauginos}.
As it was first noticed in Ref.~\cite{Fox:2002bu},
the gauge loop of  fig.~\ref{fig:feyndiagDivergence0}
gives a   finite contribution to  the scalar soft masses,
as can be  seen by simple power counting of this diagram.
One obtains \cite{Fox:2002bu}
\be
\label{gauginocon}
m_i^2=\sum_a \frac{C^i_a g^2_a M^2_{a}}{4\pi^2}\ln\frac{M_{\Phi_a}^2}{M^2_a}\, ,
\ee
where for a scalar $i$ in the fundamental representation
of $SU(3)_c\times SU(2)_L\times U(1)_Y$
we have  $C^i_a=(4/3,3/4, Y_i^2)$, while  $g_a$ are the gauge couplings of group $a$,
$M_{a}$  the gaugino masses
and $M_{\Phi_a}$  the supersymmetry-breaking masses of the real part of the scalar component of $\Phi_a$.

On the other hand, squark masses arising from \eq{topmass}, as illustrated in  fig.~\ref{fig:feyndiagDivergence},
are quadratically divergent.  
The contribution to stop soft masses is
\be
\label{yukacon}
m_{U}^2= 2 m^2_Q\simeq \frac{Y_t^2}{8\pi^2}\Lambda^2\, ,
\ee 
where we have identified the  momentum cut-off  with $\Lambda\sim$ TeV,  the scale at which 
the  operator \eq{topmass}  is induced.
Interestingly, the equivalent one-loop  contribution  for  the Higgs soft-mass,
 the first  diagram  of fig.~\ref{fig:feyndiagDivergence2},  vanishes.
 This can be understood as follows.
If we are  interested  only in  the  scalar component of $H$, we can neglect  the $\theta$-dependent part of $H$
and write the top Yukawa coupling as  $\int d^2\theta\ Y_u H^\dagger Q U=Y_uH^\dagger \int d^2\theta\,  Q U$ that is  supersymmetric
and then cannot  generate  soft-breaking terms.
At the two-loop level,   however, where the full Higgs superfield $H$  can propagate (see fig.~\ref{fig:feyndiagDivergence2}),
we do  expect  a nonzero Higgs soft-mass to be induced.
Surprisingly, we   find that the contribution
 arising from the second diagram of  fig.~\ref{fig:feyndiagDivergence2}  vanishes, and only
     the third diagram    induces   a nonzero $m_H^2$.
The latter is proportional to the squark masses, and,  as in the MSSM,
diverges  logarithmically:
  \be
  m^2_H\simeq-\frac{3Y_t^2}{16\pi^2}\left[m^2_Q\ln\frac{\Lambda^2}{m^2_Q}+
  m^2_U\ln\frac{\Lambda^2}{m^2_U}\right]\, .
  \ee
There is also a  contribution to the Higgs soft-mass  arising from 
\eq{extraquartic} (see fig.~\ref{fig:feyndiagDivergence3})
that  diverges quadratically: 
\be
m_H^2\simeq \frac{3\delta \lambda}{2\pi^2}\Lambda^2\, .
\label{quarcon}
\ee 
We can then conclude that the natural values for the  stop  masses  are 
\be
m_{Q,U}^2\simeq (400\ {\rm GeV})^2\Bigg[\left(\frac{M_{\tilde g}}{2\ {\rm TeV}}\right)^2
\ln\frac{M_{\Phi_{\tilde g}}^2}{M^2_{\tilde g}}
+(0.15,0.3)\left(\frac{\Lambda}{2\ {\rm TeV}}\right)^2\Bigg]\, ,
\ee 
where we have used \eq{gauginocon} and \eq{yukacon}.
For the Higgs soft-mass we expect
\bea
m_H^2\simeq -(100\ {\rm GeV})^2&\Bigg[
1.9\left(\frac{m_{Q}}{400\ {\rm GeV}}\right)^2
\frac{\ln\frac{\Lambda}{m_Q}}{\ln 5}
-3.2\left(\frac{M_{\tilde W}}{2\ {\rm TeV}}\right)^2
\ln\frac{M_{\Phi_{\tilde W}}^2}{M^2_{\tilde W}}
-\left(\frac{\delta \lambda}{0.015}\right)\left(\frac{\Lambda}{2\ {\rm TeV}}\right)^2
\Bigg]\, ,
\eea
where we have used \eq{gauginocon}, \eq{yukacon} and \eq{quarcon}, and  taken $m_U\sim m_Q$.
This shows that   EWSB can occur naturally at $\langle H\rangle=v\simeq 174$ GeV  without a major tuning of parameters
for  $\Lambda$  and gaugino masses around 2 TeV, and   stops and left-handed sbottoms  around 400 GeV.

The rest of the scalars   are expected also to get  masses from at least the gaugino loops (fig.~\ref{fig:feyndiagDivergence0}), although 
they could also   have couplings of order one to $X/M$ such that their masses   would then be of order TeV.
As it is well known, this does not create naturalness  problems \cite{Dimopoulos:1995mi}.
 This scenario would really correspond to the most minimal low-energy supersymmetric model   with only the stops/sbottoms 
 and  (possibly)  the gravitino below the TeV scale.

%%%%%%%%%%%%%%%%%%%%%%%%%%%%%%%%%%%%%%%%
%%%%%%%%%%%%%%%%%%%%%%%%%%%%%%%%%%%%%%%%%%%%%%%%%%%%%%%%%%%%%%%%%%%%%%%%%%%%%%%%%%%%%%%%%%%%%%%%%%%%%%%%%%%%%%%%%%%%%%%%
\section{Phenomenological Implications}

%%%%%%%%%%%%%%%%%%%%%%%%%%%%%%%%%%%%%%%%
%%%%%%%%%%%%%%%%%%%%%%%%%%%%%%%%%%%%%%%%
\subsection{The 125 GeV Higgs}\label{Higgspheno125}

\noindent

Differently from the MSSM, this supersymmetric model possesses only one Higgs scalar, identified with a neutrino superpartner, while the charged scalars in the same isospin multiplet are the Nambu-Goldstone bosons responsible for $W^\pm$ and $Z$ masses.  
At the renormalizable level,  the Higgs couplings to the SM fermions and gauge bosons 
are the same as those of the SM Higgs,
and deviations  can only arise from loop effects or higher-dimensional operators.
Potentially, the  most important   effects on the Higgs phenomenology   come from 
\emph{i)} loops  mediated by  light stops, 
\emph{ii)}  invisible decay  into  neutrino $\nu_L$ and  gravitino, 
and 
 \emph{iii)} Higgs coupling modifications from higher-dimensional operators. \\

\begin{figure}[htbp]
\begin{center}
\includegraphics[width=0.6\textwidth]{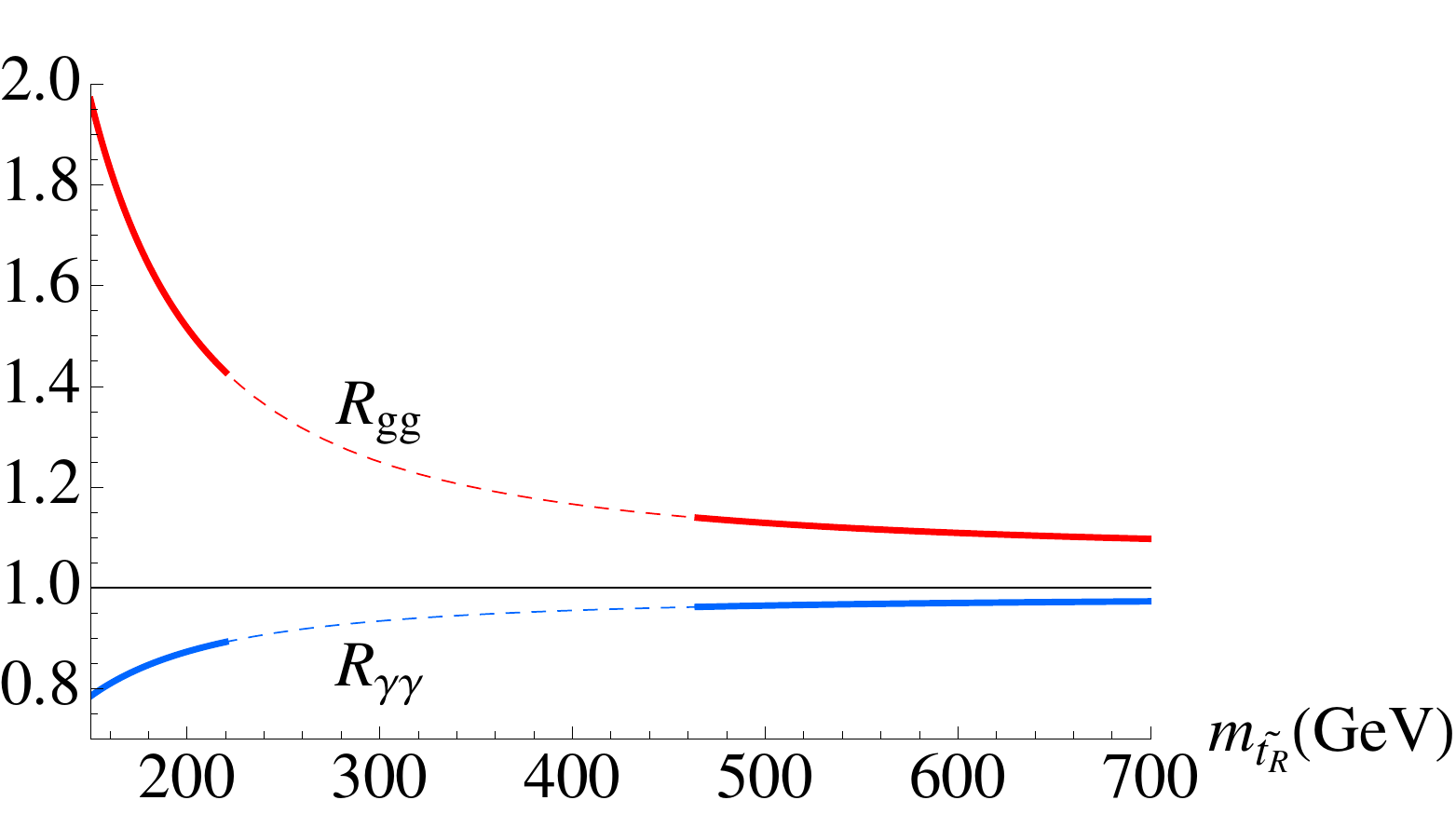}
\caption{\emph{The ratio $R=\Gamma/\Gamma^{SM}$ for the partial width of $h\to gg$  and $h\to\gamma\gamma$  as a function of $m_{\tilde{t}_R}$ while keeping  $m_{\tilde{t}_L}=500\GeV$. The dashed part corresponds to a region that is already excluded by direct searches \cite{LightStops,Stops} (see later). }}
\label{Fig:Rgg}
\end{center}
\end{figure}

\noindent 
 \emph{i)} The only  scalars  that can give sizable modifications to the Higgs couplings are the stops.
 Other scalars, even if light,  have  a small  impact on  Higgs physics 
 since their couplings to the Higgs are small.
On the contrary,  light stops  can  give sizable loop contributions to the  effective Higgs couplings to gluons and photons. 
The Higgs decay width to photons is corrected as  \cite{Djouadi:2005gj}
\begin{equation}
R_{\gamma\gamma}\equiv\frac{\Gamma_{h\gamma\gamma}}{\Gamma^{SM}_{h\gamma\gamma}}\simeq \left|1-0.2\sum_{L,R}\frac{D_{\tilde{t}_{L,R}}+m_t^2}{m^2_{\,\tilde{t}_{L,R}}}A_{0}(\tau_{\,\tilde{t}_{L,R}})\right|^2\, ,
\label{coup}
\end{equation}
where 
 $D_{\tilde{t}_L}\equiv (1/2-2s_W^2/3)m_Z^2$, $D_{\tilde{t}_R}\equiv(2s_W^2/3)m_Z^2$,  $\tau_{\,\tilde{t}_{L,R}}\equiv m_h^2/(4m_{\,\tilde{t}_{L,R}}^2)$ and, in the region of parameter space that we consider here, $A_{0}(\tau)\equiv \tau^{-2}(\arcsin^2\sqrt{\tau}-\tau)$, which has the limit  $A_0(\tau\to0)=\frac{1}{3}$. 
Similarly, the effective coupling to gluons, and hence the production cross-section, is modified as,
\begin{equation}\label{RGG}
R_{gg}\equiv\frac{\sigma_{hgg}}{\sigma^{SM}_{hgg}}\simeq \left| 1+0.7\sum_{L,R}\frac{D_{\tilde{t}_{L,R}}+m_t^2}{m^2_{\,\tilde{t}_{L,R}}}A_{0}(\tau_{\,\tilde{t}_{L,R}})\right|^2,
\end{equation}
where the same formula holds for the decay width $\Gamma_{hgg}$.
Notice that the effects of light scalars on $\Gamma_{h\gamma\gamma}$ are generally small as compared with the SM loop contribution (which includes  $W^\pm$), while the effects on the production cross-section can be sizable. We show this in fig.~\ref{Fig:Rgg} by plotting the ratio of the width for $h\to gg,(\gamma\gamma)$  in our model as 
compared with the SM.\\

\noindent 
 \emph{ii)}  A genuine   property
 of models  in which  the Higgs and neutrino are superpartners is
their interaction  with the goldstino,  that is fixed by supersymmetry to be
 \begin{equation}\label{gravcoup}
\mathcal{L}= \frac{1}{\sqrt{2}F}\partial_\mu h\,\, \partial_\rho\tilde{G}\,\sigma^\mu\bar{\sigma}^\rho\nu_L+\textrm{h.c.}\, .
\end{equation}
If the gravitino is light  this  coupling  induces  invisible Higgs decays into neutrino and gravitino:~\footnote{This possibility has been already considered in ref.~\cite{Antoniadis:2004se} in the context of non-linearly realized supersymmetry as arising from specific string constructions.}
\begin{equation}
\Gamma({h\rightarrow  \tilde{G}\nu_L})\simeq\frac{1}{16\pi}\frac{m_h^5}{F^2},
\end{equation}
where we have neglected the small masses  of the final states. For a 125 GeV Higgs, this invisible width equals the decay width into $b\bar{b}$ for $F\simeq (700\GeV)^2$, while for $F\simeq (1\TeV)^2$ it  induces an invisible fraction of about  $10\%$. 
Therefore  this  invisible Higgs decay  can be
a   striking feature   of  this supersymmetric  scenario  if supersymmetry  is broken at the TeV.   
Invisible Higgs decays, however, are 
also   present  in other well-motivated scenarios such as, for instance, 
composite  Higgs models in which 
  the Higgs can   decay    to    a composite dark matter~\cite{Frigerio:2012uc}.

\vspace{0.5cm}
\noindent 
 \emph{iii)}  Modifications to  Higgs couplings can also  arise  from higher-dimensional operators 
 induced  either from integrating out heavy superpartners 
 or from  the new physics at the scale $\Lambda$ responsible for \eq{topmass} and \eq{extraquartic}.
 Of the first type, only those from integrating the wino can give a tree-level
correction to the Higgs coupling to $l_L$, but this is quite small,
of order $g^2 v^2/M^2_{\tilde W}\lesssim 0.01$.
Corrections from  higher-dimensional operators suppressed  by $\Lambda$ 
can give effects of order
 $g_M^2v^2/\Lambda^2$
where  $g_M$  generically denotes  the coupling of the Higgs to the new sector at $\Lambda$, and therefore
can be  larger if $g_M>1$. 
The only higher-dimensional supersymmetry-preserving operator that can be written is \cite{Antoniadis:2009rn}
\begin{equation}
g_M^2 \int d^4\theta \frac{(H^\dagger e^V H)^2}{\Lambda^2},
\label{custo}
\end{equation}
where $V$ denotes the SM vector superfields.
This operator however contributes also to the $T$-parameter, which is strongly constrained by precision tests
 \cite{Giudice:2007fh}, requiring
$g_Mv/\Lambda\lesssim 10^{-3}$ and therefore small corrections to the Higgs couplings~\footnote{In strongly-interacting 
Higgs models   \cite{Giudice:2007fh}
a custodial $SU(2)$ symmetry,  under which  $(H,H^c)$ transforms as a $\bf 2$,  
can be implemented to avoid large corrections to $T$.
Nevertheless,  this custodial symmetry
does not commute with supersymmetry and thus cannot be used here to protect the $T$-parameter.}.
It is important to notice that   \eq{custo} can only be  generated  at tree-level from integrating out     heavy singlets, and then
it is not  generated in models in which  only heavy doublets are present.
In the Appendix we propose a simple UV completion of our model
which involves  extra heavy Higgs superfields. In this case, certain corrections to the Higgs 
couplings  can be sizable   (see \eq{topmodi})  
 without conflicting  with the $T$-parameter. 
\begin{figure}[top]
\begin{center}
\includegraphics[height=0.375\textwidth]{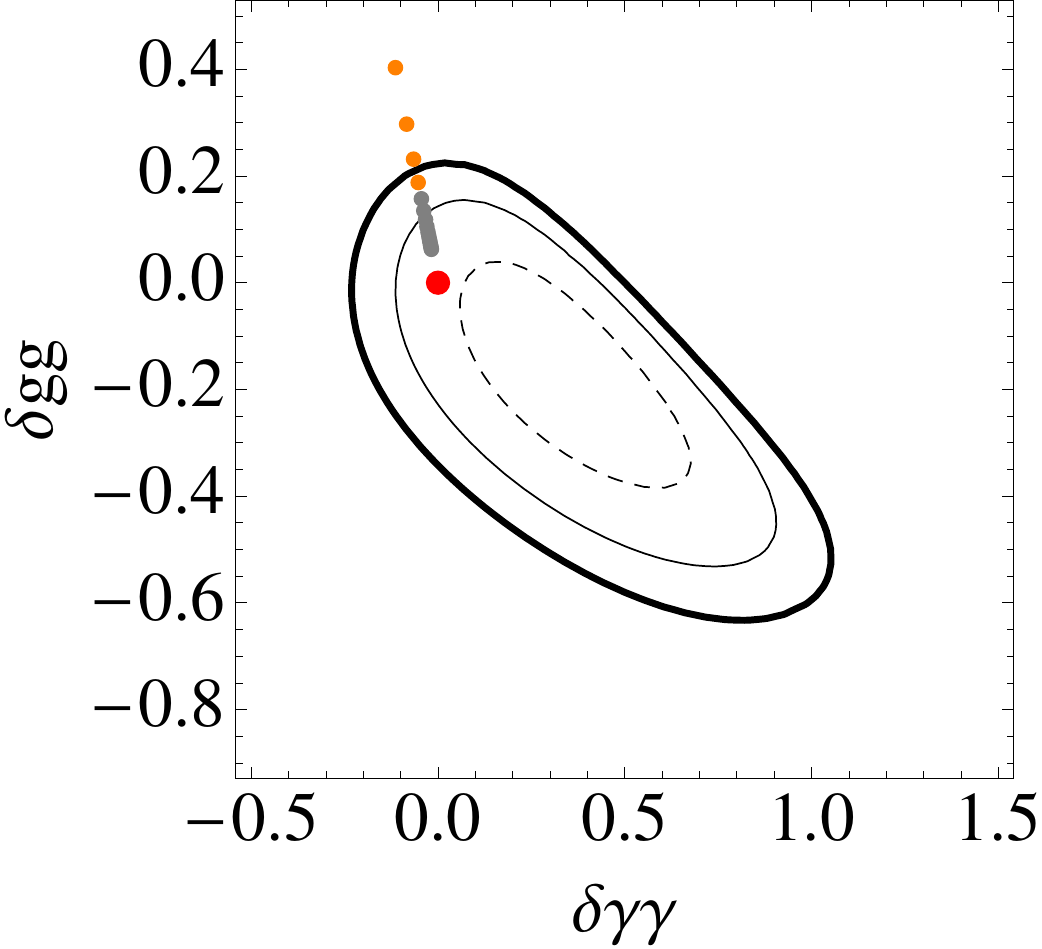}
\hspace{.8cm}
\includegraphics[height=0.4\textwidth]{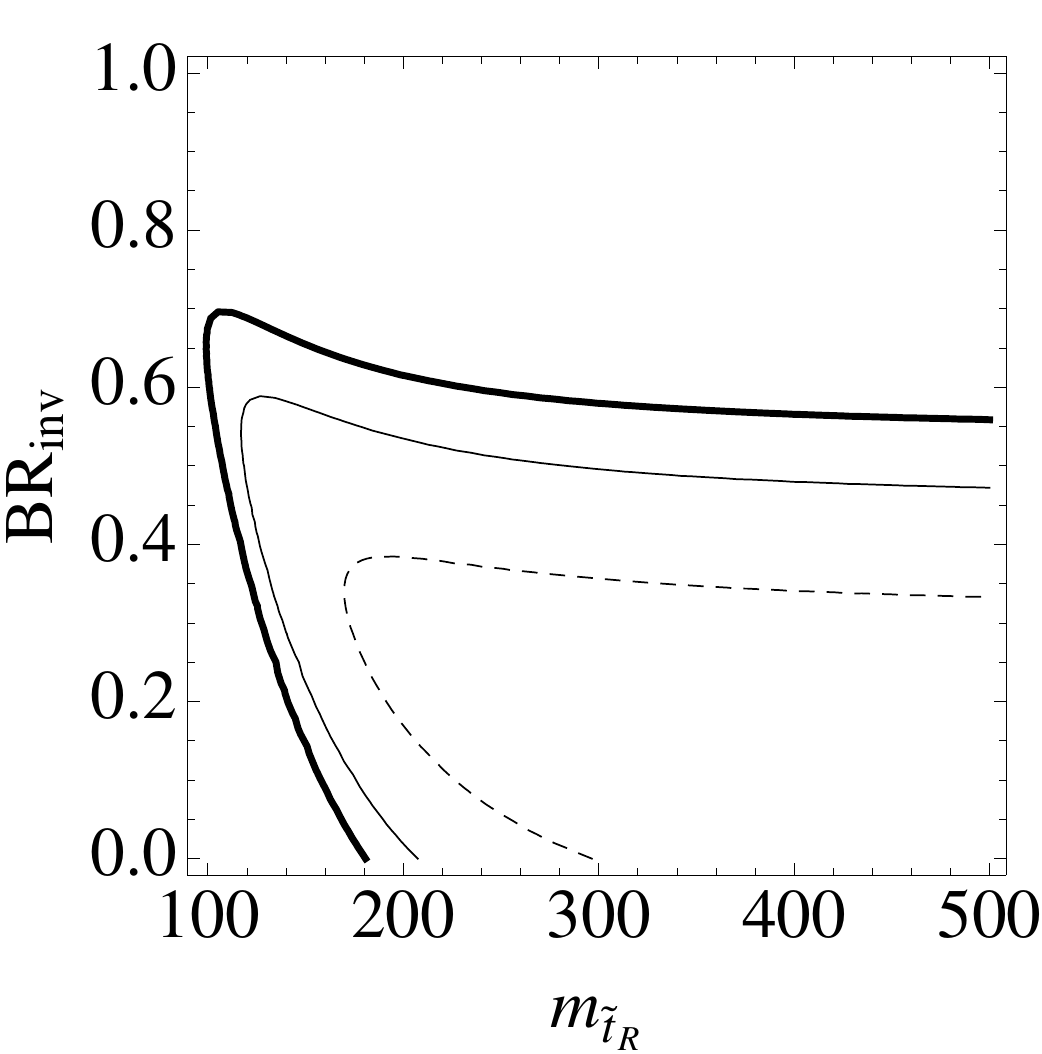}
\caption{\emph{Left panel:  the $68\%,95\%$ and $99\%$ C.L. contours (dashed, thin, thick) on the plane $\delta_{gg}$ vs $\delta_{\gamma\gamma}$. The red point corresponds to the SM, the orange ones correspond to a light stop $\tilde{t}_R$ (from 150~GeV to 300~GeV in steps of 25 GeV) and the grey ones to a heavier stop $\tilde{t}_R$ (up to 500~GeV). Right panel: same contours on the plane $BR_{inv}$ vs $m_{\tilde{t}_R}$.}}
\label{Fig:bestfit}
\end{center}
\end{figure}

In summary, 
the most important effects
that characterize a Higgs as a superpartner
of the neutrino  are 
 its   invisible decay to neutrino and gravitino   and   possible  
modifications  of  the effective $hgg(\gamma\gamma)$ couplings if the stops are light.
It is interesting to compare these predictions with the  
experimental data   recently extracted for  the   125 GeV Higgs.
This is done 
 in fig.~\ref{Fig:bestfit}  where we show 
  the $68\%,95\%$ and $99\%$ C.L.  contours  obtained
after performing    a  $\chi^2$ analysis of the recent experimental data following  ref.~\cite{Montull:2012ik}.
 The lefthand panel shows the preferred regions for the parameters $\delta_{gg,\gamma\gamma}$ defined as the deviations from the SM effective couplings between the Higgs and the gluons/photons, $R_{gg,\gamma\gamma}=(1+\delta_{gg,\gamma\gamma})^2$. 
The theoretical prediction for our model, as extracted from \eq{coup} and \eq{RGG} for different $\tilde{t}_R$ masses, is also shown.  
 Noticably, the impact of a light stop is to worsen the Higgs coupling fit. Nevertheless, the presence of a nonzero $BR_{inv}$ tends to improve the fit and for  $BR_{inv}\gtrsim 0.2$  the fit can be comparable with the SM even for light stops, as we show in the righthand panel where we plot  the preferred regions in the parameter space of our model ($m_{\tilde{t}_R},BR_{inv}$).
 In both plots we have kept $m_{\tilde t_L}=530$ GeV since, as we will see later, this is the experimental lower-bound.
Although present experimental data is  not  decisive, future data should be able to  favor or disfavor this scenario.

%%%%%%%%%%%%%%%%%%%%%%%%%%%%%%%%%%%%%%%%%%%%%%%%%%%%%%%%%%%%%%%%%%%%%%%%%%%%%%%%%%%%%%%%%%%%%%%%%%%%%%%%%%%%%%%%%%%%%%%%%%%%%%%%%%%%%%%%%%%%%%%%%%%%%%%%%%
\subsection{Stops and sbottoms}

Models  in which the  Higgs  is the neutrino superpartner  have   
a   squark phenomenology   different from the ordinary MSSM.
We  focus first  on  the third generation squarks  which naturalness arguments
suggest to be the lightest.

Since  the $U(1)_R$ symmetry  forbids  supersymmetry-breaking  trilinear $A$-terms,
the left-handed and right-handed squarks  do not mix  
and  are mass eigenstates. 
One important consequence   is that $\tilde{b}_L$ is always lighter than  $\tilde{t}_L$, since their 
masses are related by
\begin{equation}\label{sbottomstop}
m^2_{\tilde{b}_L}=m^2_{\tilde{t}_L}-m^2_{t}+m^2_{b}.
\end{equation}
The possible decay  modes of the squarks  are dictated by symmetries. 
One can easily see that  Lorentz, electromagnetic and $U(1)_R$  symmetry  only allow the decay channels
shown   in table~\ref{tab:decays}. 
\begin{table}
\begin{minipage}[b]{0.5\linewidth}\centering
\begin{tabular}{|l|l|}
\hline
Decay  & Interaction \\
\hline
\hline
$\tilde{t}_L \to b_R \bar l^-_L $  & $Y_d\, HQD|_{\theta^2}$  \\
\hline
$\tilde{t}_L \to t_R \bar{\nu}_L $ &$\frac{1}{\Lambda^2}|H|^2|Q|^2|_{\theta^4}$ \\
\hline
$\tilde{t}_L \to t_L \tilde{G} $ &$\frac{m_{t}^2-m_{\tilde{t}_L}^2}{F}\,\tilde{t}_L^*\tilde{G}\,t_L$ \\
\hline
\hline
$\tilde{b}_L \to b_R \bar{\nu}_L $  & $Y_d\, QHD|_{\theta^2}$ \\
\hline
$\tilde{b}_L \to b_L  \tilde{G}  $   &  $ \frac{m_{b}^2-m_{\tilde{b}_L}^2}{F}\,\tilde{b}_L^*\tilde{G}\,b_L$ \\
\hline
\end{tabular}
\end{minipage}
\begin{minipage}[b]{0.5\linewidth}
\centering
\begin{tabular}{|l|l|}
\hline
Decay  & Interaction \\
\hline
$\tilde{t}_R \to t_L \nu_L $ & $\frac{1}{\Lambda^2}|H|^2|U|^2|_{\theta^4}$  \vspace{0.5mm} \\
\hline
$\tilde{t}_R \to t_R \bar{\tilde{G}} $ & $ \frac{m_{t}^2-m_{\tilde{t}_R}^2}{F}\,\tilde{t}_R^*\bar{\tilde{G}}\,\bar{t}_L$\\
\hline
\hline
$\tilde{b}_R \to b_L \nu_L $  & $Y_d\, QHD|_{\theta^2}$  \\
\hline
$\tilde{b}_R \to t_L \, l^-_L $  & $Y_d\, QHD|_{\theta^2}$  \\
\hline
$\tilde{b}_R \to b_R  \bar{\tilde{G}} $   &  $ \frac{m_{b}^2-m_{\tilde{b}_R}^2}{F}\tilde{b}_R^*\bar{\tilde{G}}\,\bar{b}_L$ \\
\hline
\end{tabular}
\end{minipage}
\caption{\emph{Decay modes for the (third family) squarks
with the corresponding Lagrangian interaction.}}
\label{tab:decays}
\end{table}
These decays  can arise from the following interactions.
From the 
superpotential term $Y_b\, HQD$ in \eq{W}, we have contributions 
to
\begin{equation}\label{leptonicdecays}
 \tilde{t}_L \to b_R \,\bar l^-_L ,\quad \tilde{b}_L \to b_R \bar{\nu}_L,\quad \textrm{and}\quad \tilde{b}_R \to b_L \nu_L,\, t_L l^-_L.
\end{equation}
Goldstino interactions, as in the MSSM, arise from
\begin{equation}\label{Gravitino}
\frac{1}{F}\partial_\mu \tilde{t}^*_{L}\,\, \partial_\rho\tilde{G}\,\sigma^\mu\bar{\sigma}^\rho t_{L}\,\,
 \stackrel{\textrm{on-shell}}{=} \,\,\frac{(m_t^2-m_{\tilde{t}_{L}}^2)}{F}\,\tilde{t}^*_{L}\,\tilde{G}\,\, t_{L}\, ,
\end{equation}
and similarly for other squarks, that leads to 
\begin{equation}
\label{eq:goldzzz}
\tilde{t}_R \to t_R \bar{\tilde{G}},\quad \tilde{t}_L \to t_L \tilde{G},\quad 
\tilde{b}_R \to b_R \bar{\tilde{G}},\quad \tilde{b}_L \to b_L \tilde{G}\, .
\end{equation}
Exchanges of heavy winos and binos  leads to effective interactions between (s)quarks and leptons, such as
\begin{equation}\label{effectivegaugino}
\frac{2 g'^2v}{3 M_{\tilde{B}}^2}\tilde{t}_R \bar{t}_R\deslash \nu_L\, ,\quad 
\frac{ g^2v}{2 M_{\tilde{W}}^2}\tilde{t}_L \bar{b}_L\deslash l^-_L\, .
\end{equation}
However, due to the Dirac nature of the gauginos, the structure of these interactions is such that the decay amplitudes are proportional to the final-state lepton mass $\sim v m_{\nu,l}/M_{\tilde{W}}^2$ and are therefore  very small.  
Such decays, however,
could also arise from  dimension-six operators that might be induced at the scale $\Lambda$.
For example
\begin{equation}
\int d^4\theta\frac{1}{\Lambda^2}|H|^2|Q|^2\  ,\quad \int d^4\theta\frac{1}{\Lambda^2}|H|^2|U|^2\, ,
\end{equation}
induce
\begin{equation}
\tilde{t}_{R}\to t_L \nu_L,\quad
\tilde{t}_{L}\to t_R \bar{\nu}_L\, ,
\label{sqneu}
\end{equation}
with an amplitude  proportional to the top mass.
Note that in these decays, in the limit $m_t\ll m_{\tilde t_{L,R}}$,  the top helicity  is  fixed:  $U(1)_R$ charge conservation requires, for $\tilde{t}_R$, a top and a neutrino (rather than an anti-neutrino) in the final state, while spin conservation implies that, in the stop rest frame, the quark helicity be opposite to  the neutrino helicity; and vice versa for $\tilde{t}_L$. 
This offers an interesting way to differentiate between the squarks decays of \eq{sqneu} and those of 
\eq{eq:goldzzz} that are also present in the MSSM with low-scale supersymmetry breaking,
since these latter produce final-state tops with opposite helicity.

\begin{figure}[top]
\begin{center}
\includegraphics[width=0.5\textwidth]{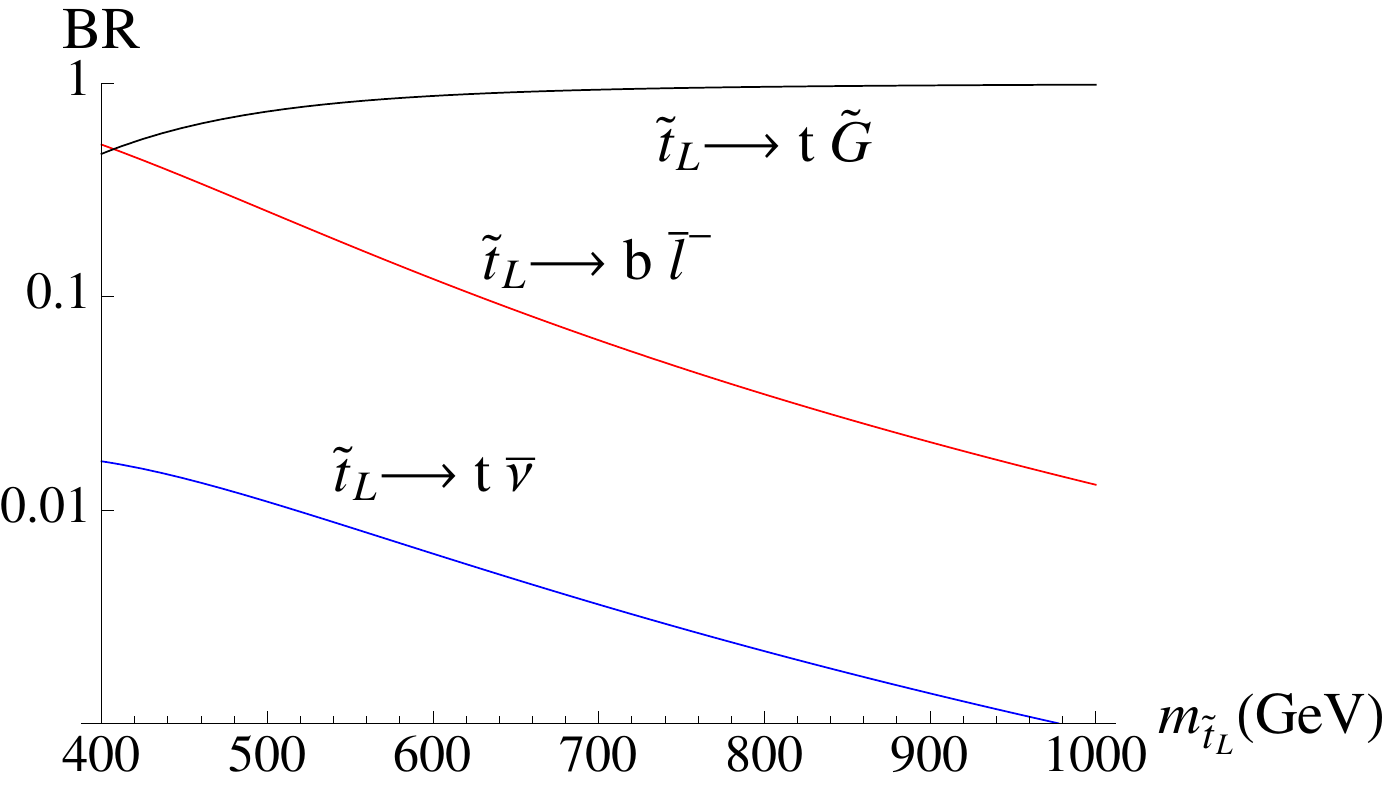}\hspace{.3cm}
\caption{Branching ratios for  $\tilde{t}_L$ decays as a function of its mass, for $\Lambda=\sqrt{F}=2$~TeV.}
\label{fig:Br}
\end{center}
\end{figure}

Let us now discuss the size of the different branching ratios for stops and sbottoms.
In fig.~\ref{fig:Br} we compare the branching ratios of $\tilde{t}_L$ into different channels for 
$\Lambda=\sqrt{F}=2$~TeV.
We can see that  the decays into gravitinos dominate, but 
the branching ratio  into  
 $b$ and leptons is sizable enough  to allow detection.
For larger values of $F$, or in models in which  the gravitino is heavy (such as models with emergent supersymmetry), 
$\tilde{t}_L$    can decay  dominantly into $b +\bar l^-$. 
Indeed, for $\sqrt{F}\gg $ TeV,  the ratio  between the two dominant stop decay widths  is given by
\begin{equation}\label{BRStauL}
\frac{\Gamma(\tilde{t}_L\to b \, \bar l^-)}{\Gamma(\tilde{t}_L \to t \bar{\nu})}\simeq \frac{m_b^2}{m_t^2}\frac{\Lambda^4}{v^4}\left(1-\frac{m_t^2}{m_{\tilde{t}_L}^2}\right)^{-2}\simeq 10\left(\frac{\Lambda}{2\TeV}\right)^4\, ,
\end{equation}
showing that for $\Lambda\gtrsim 1 \TeV$ the decay into $b+\bar l^-$
dominates. 

For  $\tilde b_R$, on the other hand,  the branching ratios into $b_L\nu_L$ and $t_L l_L^-$ are comparable, as both are controlled by the Yukawa $Y_b$,
\begin{equation}\label{BRSbottom}
\frac{\Gamma(\tilde{b}_R \to t_L l_L^-)}{\Gamma(\tilde{b}_R\to b_L \, \nu_L)}\simeq \left(1-\frac{m_t^2}{m_{\tilde{b}_R}^2}\right)^{2}\, .
\end{equation}
Nevertheless, for small  $F$ and  a light gravitino,  the decays into gravitinos  dominate:
\begin{equation}\label{sbottomgravitino}
\frac{\Gamma(\tilde{b}_R \to  b  \tilde{G} )}{\Gamma(\tilde{b}_R\to b_L \, \nu_L)}\simeq \frac{m_{\tilde{b}_R}^4}{F^2}\frac{v^2}{m_b^2}\simeq 7 \left(\frac{m_{\tilde{b}_R}}{500 \GeV}\right)^{4} \left(\frac{2 \TeV}{\sqrt{F}}\right)^{4}\, .
\end{equation}
This same expression \eq{sbottomgravitino} holds for  $\tilde{b}_L$,   
for which the decay  
into charged leptons is forbidden by symmetries.
Finally, for  $\tilde t_R$ we find
\begin{equation}\label{stopRgravitino}
\frac{\Gamma(\tilde{t}_R \to t_R \bar{\tilde{G}} )}{\Gamma(\tilde{t}_R \to t_L \nu_L)}\simeq \frac{m_{\tilde{t}_R}^4}{v^2m_t^2}\frac{\Lambda^4}{F^2}\simeq 70 \left(\frac{m_{\tilde{t}_R}}{500 \GeV}\right)^{4} \left(\frac{\Lambda^2}{F}\right)^2\, .
\end{equation}

\begin{figure}[htbp]
\begin{center}
\includegraphics[width=0.6\textwidth]{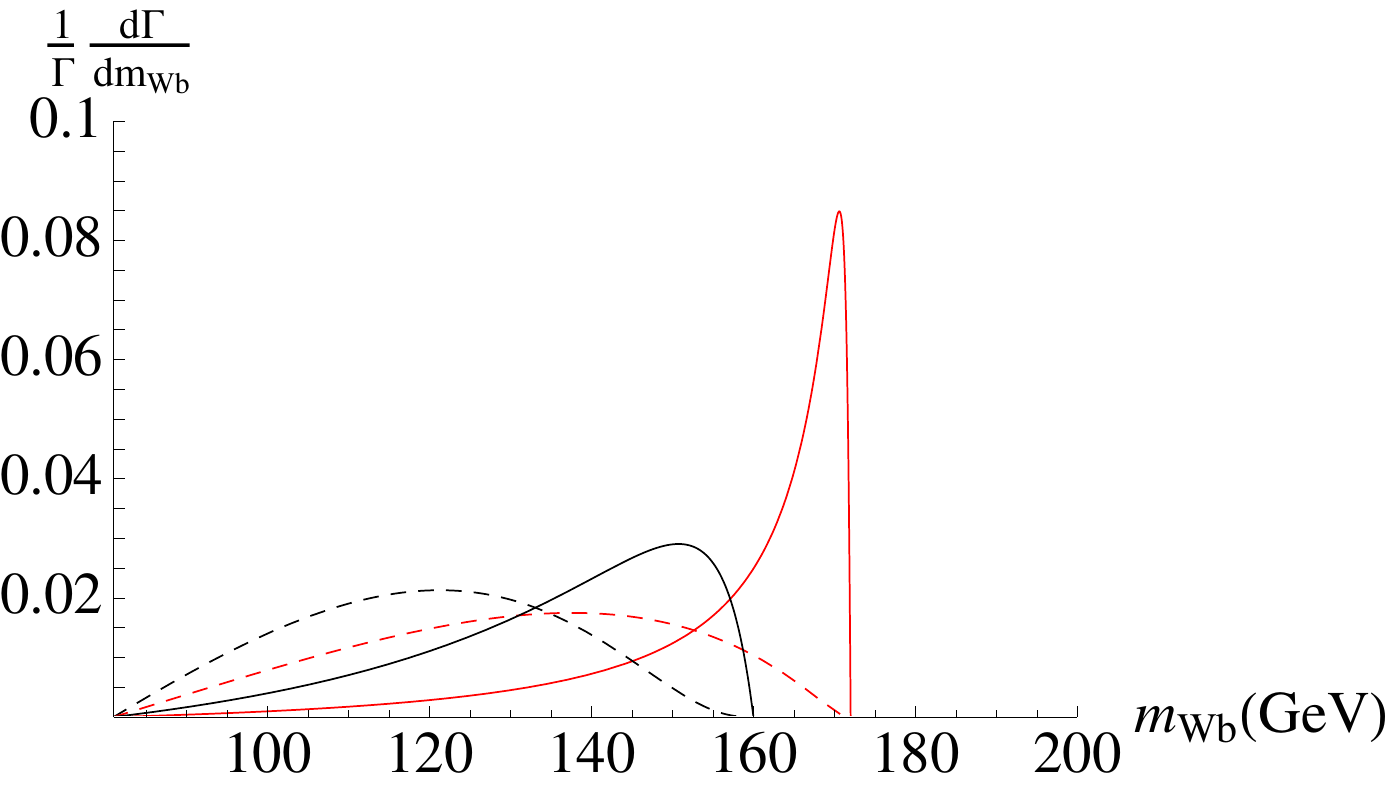}
\caption{\emph{Invariant mass distribution of the $W$ and $b$ in the decay $\tilde{t}_R\rightarrow W\,b\,\tilde{G}$ (dashed curves) and $\tilde{t}_R\rightarrow W\,b\, \nu_L$ (solid curves) for $m_{\tilde{t}_R}=160\, (172)\GeV$ in black (red), taking $m_t=173\GeV$.}}
\label{FigDist}
\end{center}
\end{figure}

\vspace{1cm}
\noindent
\emph{Searches}\\
As discussed above, many decay processes  have neutrinos or gravitinos in the final state, resulting in signatures with missing energy, which resemble much those  of the MSSM. For this reason we can adapt present LHC searches to our model. This is particularly true for $\tilde{b}_L$ 
whose decay  final state is always a  bottom-quark plus missing energy. This has the same signature as the MSSM decay into bottom plus neutralino, in the limit where the neutralino is massless, and is presently searched for at the LHC \cite{Atlassbottom,SbottomCMS}. 
Present exclusion bounds amount to
\begin{equation}\label{sbottommassbound}
m_{\tilde{b}_L}> 500\GeV\, .
\end{equation}
From \eq{sbottomstop}, bounds on the sbottom mass imply a bound on the $\tilde t_L$ mass:
\begin{equation}
m_{\tilde{t}_L}\gtrsim 530\GeV\, .
\end{equation}
Similarly, searches for $t$ and missing energy, motivated by the MSSM decay pattern  $\tilde{t}\to t \chi_0$ with a massless neutralino, also cover $\tilde{t}_R$ decays in our model. 
The mass range \begin{equation}
220\GeV\lesssim m_{\tilde{t}_R}\lesssim 465\GeV\, ,
\end{equation}
is already excluded by a combination of searches \cite{Stops300,Stops3002,Stops}. Searches for stops lighter than the top (or almost degenerate)   are reputedly very hard \cite{Kats:2011it,Kats:2011qh,Bai:2012gs} and, to our knowledge, the best bound that can be extrapolated gives \cite{Kats:2011it,Kats:2011qh}
\begin{equation}\label{lightstopRBound}
m_{\tilde{t}_R} \gtrsim 150\GeV\, ,
\end{equation}
in the low-mass range.
As commented above, it can be possible to distinguish between $\tilde{t}_R$ decays into gravitinos or into neutrinos by measuring the helicity of the final state tops. This is feasible if $m_{\tilde{t}_R}\gg m_t$. Indeed, in this case the final-state tops are boosted (so boosted that helicity almost coincides with chirality), and they decay before hadronization so that the distribution of its decay products can be measured and the helicity extracted \cite{Krohn:2009wm}. Another interesting feature that singles out this model is that, for $m_{\tilde{t}_R}\lesssim m_t$ (a region not yet excluded by direct searches, cf. \eq{lightstopRBound}), the distribution in momentum of the decay products of the top quark in the decay $\tilde{t}_R\to t \tilde{G}$ is different from the distribution in the $\tilde{t}_R\to t \nu_L$ decay, due to the derivatives in the gravitino interaction \eq{Gravitino}; this is illustrated in fig.~\ref{FigDist}.

The supersymmetry searches described above also cover $\tilde{t}_L$
and $\tilde{b}_R$ when their dominant decays are into  quarks plus
neutrinos or gravitinos. On the contrary, for $\sqrt{F}\gtrsim few
\TeV$ or if the gravitino is heavy, $\tilde{t}_L$ and $\tilde{b}_R$
have sizable  decay widths into bottom/top quarks and charged leptons, see
Eqs.~(\ref{BRStauL}), (\ref{BRSbottom}) and fig.~\ref{fig:Br}. In this case, also LHC searches for leptoquarks apply to our model and, depending on the  flavor of $l_L$, the present bounds for $\tilde{t}_L$ are~\footnote{Dedicated searches for leptoquarks $\to$ $b$-jets+$\mu/e$ with $b$-tagging could improve the sensitivity to our model for $l^-_L=e_L,\mu_L$.}
\begin{eqnarray}\label{stopleftmassbound}
m_{\tilde{t}_L}>\left\{\begin{array}{ccr}660\GeV & l^-_L=e_L &\textrm{\cite{Aad:2011ch}} \\685\GeV & l^-_L=\mu_L&\textrm{\cite{Aad:2012cy}} \\  525\GeV & l^-_L=\tau_L&\textrm{\cite{LeptoquarksCMS}}\end{array}\right. ,
\end{eqnarray}
while for $\tilde{b}_R$, which decays with 50\%  probability into $b\nu_L$ and  $t l^-_L$ (see \eq{BRSbottom}), the best bounds come from searches on the decay product  $b\nu_L$ ($b$-quarks plus missing energy \cite{Atlassbottom,SbottomCMS})  that  lead
to  $m_{\tilde{b}_R}> 500\GeV$.

Due to the lack of  ordinary MSSM $R$-parity, $\tilde{t}_L$ and $\tilde{b}_R$ squarks can be singly produced in this model. Nevertheless,  the production cross-sections for these processes  
are  proportional to $Y^2_b$  and are then  very small ($\ll \textrm{fb}$ at 14 TeV). Furthermore, their topology (with a final state including $t$+$l_L^-$+$b$-jet and missing energy) coincides with that of  a double produced squarks when the two squarks decay differently. This leaves little hope to single out this feature in the early phases of LHC .

We conclude with a possible   strategy to differentiate between  third family squarks of ordinary supersymmetric models and of models where  the Higgs is a neutrino superpartner. If a scalar resonance decaying into $b$-jets and missing energy is observed, it can be our $\tilde{b}_R$ only if also leptoquark decays are observed at the same mass. If no leptoquark decays are observed then it could still be our $\tilde{b}_L$, but from \eq{sbottomstop}, this would imply that another scalar resonance, the $\tilde{t}_L$, must  be observed at slightly heavier mass. 
On the other hand,  the observation of a scalar decaying  into $t+\!\not \!\!\!E_T$ could be attributed to our $\tilde{t}_L$ if also 
decays into $b +\bar l^-_L$ are seen.  If not, there is still the possibility to   be our $\tilde{t}_R$. 
To know whether this is the case, we must discriminate 
 between the decay $\tilde{t}_R \to t_L \nu_L $, typical of our model,  and  the decay $\tilde{t}_R \to t_R \bar{\tilde{G}} $ common to many supersymmetric models. 
 For $m_{\tilde{t}_R}\gg m_t$, this can be done by  measuring  the final-state  top-quark helicity,
while  for $m_{\tilde{t}_R}\ll m_t$, we must look at  the 
  differences in the $Wb$ invariant mass distribution, fig. \ref{FigDist}.

%%%%%%%%%%%%%%%%%%%%%%%%%%%%%%%%%%%%%%%%%%%%%%%%%%%%%%%%%%%%%%%%%%%%%%%%%%%%%%%%%%%%%%%%%%%%%%%%%%%%%%%%%%%%%%%%%%%%%%%%%%%%%%%%%%%%
\subsection{First and Second Generation Squarks and Sleptons}
If the gravitino is light and $\sqrt{F}\sim$ TeV, then
the first and second generation squarks, similarly to the third generation ones,
 decay mainly into gravitinos and light quarks. Searches for jets plus missing energy address these decays \cite{Atlass12genSquarks} and the present bound is
\begin{equation}
m>760\GeV\, .
\end{equation}
On the other hand, in models where the gravitino is heavy   (or  $\sqrt{F}\gg$ TeV), 
the situation is quite different from the third-generation squark phenomenology discussed above.
The reason is that the  2-body decay into light quarks and leptons  
 are proportional to the   Yukawa couplings
of the  first   and second generation quarks that are very small.
In particular, we have
\bea
\Gamma(\tilde u_L (\tilde d_L)\to  d+\bar l^-_L (\bar{\nu}_L))&\simeq& Y_{d}^2 \frac{m_{\tilde{u}_L(\tilde{d}_L)}}{16\pi}\, , \ \ \  \ \ 
\Gamma(\tilde{d}_R\to  u+l^-_L,d+{\nu}_L)\simeq Y_{d}^2 \frac{m_{\tilde{d}_R}}{16\pi}\, ,
\\
 \Gamma(\tilde{u}_R\to u+\nu_L)&\simeq& Y_{u}^2 \frac{m_{\tilde{u}_R}}{16\pi}\frac{v^4}{\Lambda^4}\, ,
 \eea
 and similarly for the second-generation squarks. Therefore, 3-body decays can  be important or even dominate
since they are not chirality-suppressed.
For example, the Dirac-gaugino-mediated decays
 into a 3-body final state  made of 
a  quark, a lepton and a gauge/Higgs  boson,
have a partial width  given by (neglecting   final-state masses)
\bea\label{3body}
\Gamma(\tilde{q}_{i L} \to q_j +\bar l^-_L/\bar{\nu}_L+h/Z/W)&\simeq& \frac{c_{h,Z,W} }{12288\pi^3}\left[g^{\prime\, 4}\frac{Y_{\tilde{q}_{iL}}^{2}}{4}\frac{m^5_{\tilde{q}_{i L}}}{M_{\tilde{B}}^4}+ g^4c_{ij}\frac{m_{\tilde{q}_{i L}}^5}{M_{\tilde{W}}^4}\right]\, ,\\
\Gamma(\tilde{q}_{R} \to q + l^-_L/\nu_L+h/Z/W)&\simeq &Y_{\tilde{q}_R}^{2}g^{\prime\, 4}\frac{c_{h,Z,W} }{49152\pi^3}\frac{m^5_{\tilde{q}_R}}{M_{\tilde{B}}^4}\, ,
\eea
with 
$\tilde q_{iL}=\tilde u_L, \tilde d_L$,
$\tilde q_{R}=\tilde u_R, \tilde d_R$,
$q_{j}=u,d$
and $c_{ij}=(2-\delta_{ij})^2$.
We  also have $c_W=2$, $c_{h,Z}=1$, while  $Y_{\tilde{q}}$ is the squark hypercharge.
 The same formula holds for the second generation.
 %, with $\tilde q_i=\tilde c_L,\tilde c_R, \tilde s_L, \tilde s_R$. and we recall that the analog of \eq{sbottomstop} implies for first (second) generation squarks that $\tilde{u}_L (\tilde{c}_L)$ and $\tilde{d}_L (\tilde{s}_L)$ are practically degenerate.
 In table~\ref{tab:decaysLight}  we provide
 the  dominant decay mode for each of the first and second generation squarks.
 We must notice, however,
that decays into  other squark/quark pairs, if kinematically allowed, 
 could dominate over the decays of table~\ref{tab:decaysLight} (beside being enhanced by a color factor, these channels receive contributions from gluino-exchanges, which are proportional to the strong coupling).   
\begin{table}[t]
\centering
\begin{tabular}{|l||l||l|}
\hline
$\tilde{u}_L \to d +\bar l^-_L+Z $ & $\tilde{c}_L \to s +\bar l^-_L\,\,\,\,\,  (\text{for}\ m_{\tilde{c}_L}\lesssim500\GeV)$ & $\tilde{c}_R \to c+ \nu_L \,\,\,\,\, (\text{for}\ m_{\tilde{c}_R}\lesssim600\GeV)$\\\cline{1-1}
$\tilde{d}_L \to u+\bar \nu_L+W^- $& $\hspace{5.3mm} \to s+\bar l^-_L +Z$ &$\hspace{5.3mm} \to c+ l^-_L +W^+$ \\
\hline
$\tilde{u}_R \to u+ l^-_L+W^+$ &$\tilde{s}_L \to s+\bar \nu_L$$\,\,\,\,\, (\text{for}\ m_{\tilde{s}_L}\lesssim300\GeV)$&$\tilde{s}_R \to c+ l^-_L $ (50\%)\\
\cline{1-1}
$\tilde{d}_R \to d+ l^-_L+W^+ $&$\hspace{5.3mm} \to c+\bar \nu_L +W^-$&$\hspace{5.3mm} \to s+ \nu_L $  (50\%)\\
\hline
%$\tilde{d}_L \to d + \tilde{G}  $   &  $- \frac{m_{\tilde{d}_L}^2}{F}\,\tilde{d}_L^*\tilde{G}\,b_L$ \\
%\hline
\end{tabular}
\caption{\emph{Dominant decay modes for first and second family squarks
when the  gravitino is heavy or $\sqrt{F}\gg$ TeV.}}
\label{tab:decaysLight}
\end{table}
%
%
%\begin{table}
%\begin{minipage}[b]{0.5\linewidth}\centering
%\begin{tabular}{|l|l|}
%\hline
%$\tilde{u}_L \to d +\bar l^-_L+Z $  \\
%\hline
%$\tilde{d}_L \to u+\bar \nu_L+W^- $  \\
%\hline
%$\tilde{u}_R \to u+ l^-_L+W^+$ \\
%\hline
%$\tilde{d}_R \to d+ l^-_L+W^+ $\\
%\hline
%%$\tilde{d}_L \to d + \tilde{G}  $   &  $- \frac{m_{\tilde{d}_L}^2}{F}\,\tilde{d}_L^*\tilde{G}\,b_L$ \\
%%\hline
%\end{tabular}
%\end{minipage}
%\begin{minipage}[b]{0.5\linewidth}
%\centering
%\begin{tabular}{|lr|}
%\hline
%$\tilde{c}_L \to s +\bar l^-_L $&$ (m_{\tilde{c}_L}\lesssim500\GeV)$  \\
%$\hspace{5.3mm} \to s+\bar l^-_L +Z$&$(m_{\tilde{c}_L}\gtrsim500\GeV)$  \\
%\hline
%$\tilde{s}_L \to s+\bar \nu_L$&$ (m_{\tilde{s}_L}\lesssim300\GeV)$  \\
%$\hspace{5.3mm} \to c+\bar \nu_L +W^-$&$(m_{\tilde{s}_L}\gtrsim300\GeV)$  \\
%\hline
%$\tilde{c}_R \to c+ \nu_L $&$ (m_{\tilde{c}_R}\lesssim600\GeV)$  \\
%$\hspace{5.3mm} \to c+ l^-_L +W^+$&$(m_{\tilde{c}_R}\gtrsim600\GeV)$  \\
%\hline
%$\tilde{s}_R \to c+ l^-_L $& (50\%)\\
%$\hspace{5.3mm} \to s+ \nu_L $&  (50\%)\\
%\hline\end{tabular}
%\end{minipage}
%\caption{\emph{Dominant decay modes for first and second family squarks
%when the  gravitino is heavy or $\sqrt{F}\gg$ TeV.}}
%\label{tab:decaysLight}
%\end{table}

Finally, let us briefly discuss the phenomenology of the   sleptons of $L_{1,2}$
that, we recall, contain  the other two non-Higgs-superpartner leptons,
and those of $E_{1,2,3}$.
If the gravitino is light and $\sqrt{F}\sim$ TeV, the corresponding
charged sleptons  decay  into charged leptons and gravitinos, giving missing energy (this topology is searched at the LHC in the context of MSSM decays of sleptons into leptons and (massless) neutralinos \cite{sleptons}, excluding the region $m \lesssim 200\GeV$), while sneutrinos decay invisibly into neutrinos and gravitinos and can be searched for using similar strategies as for generic DM searches (monojets or dijets and missing energy).
 On the other hand, if the  gravitino is heavy,
 the analogous of \eq{3body} applies and 3-body decays can dominate.
In table \ref{tab:decaysLept}
 we show  the dominant
 decay mode of the sleptons 
 %$\tilde L_{1,2}$ and $\tilde E_{1,2,3}$
 depending on their corresponding  flavour.
\begin{table}[t!]
\centering
\begin{tabular}{|l||l||l|}
\hline
$\tilde{e}_L \to \nu_e +\bar \nu_L+W^- $&$\tilde{\mu}_L \to \nu_\mu + \bar{\nu}_L+ W^-$ &  $\tilde{\tau}_L \to \tau+\bar \nu_L$\\
\hline
$\tilde{e}_R \to e + l^-_L+W^+ $& $\tilde{\mu}_R \to \mu +\nu_L$ (50\%)&$\tilde{\tau}_R \to \tau +\nu_L$  (50\%) \\
&$\hspace{5.3mm}  \to \nu_\mu + l^-_L$  (50\%)&$\hspace{5.4mm}  \to  \nu_\tau + l^-_L$  (50\%)\\
\hline
$\tilde{\nu}_e \to e +\bar l^-_L+Z $& $\tilde{\nu}_\mu  \to \mu+ Z +\bar{l}^-_L$&$\tilde{\nu}_\tau \to \tau +\bar l^-_L$ \\
\hline
\end{tabular}
\caption{\emph{Dominant decay modes for sleptons when the  gravitino is heavy or $\sqrt{F}\gg$ TeV.
We assume  that the slepton masses are larger than 500 GeV.}}
\label{tab:decaysLept}
\end{table}

We  see that the phenomenology of squarks and sleptons is very rich in  this model and requires a dedicated study which we plan to pursue in the forthcoming future.

%%%%%%%%%%%%%%%%%%%%%%%%%%%%%%%%%%%%%%%%%%%%%%%%%%%%%%%%%%%%%%%%%%%%%%%%%%%%%%%%%%%%%%%%%%%%%%%%%%%%%%%%%%%%%%%%%%%%%%%%%%%%%%%%%%%
\section{Conclusions}

An important question, stemming from the recent LHC discovery of a resonance at 125 GeV, is  whether or not this could be the scalar superpartner of an existing fermion, hence providing the first evidence for supersymmetry.  Since
its quantum numbers coincide with those of a neutrino, we have therefore proposed a supersymmetric model in which the Higgs is identified with one of the neutrino superpartner. 
This can be realized if  lepton number is also an $R$-symmetry such that this is not broken by the  Higgs VEV.

We have shown that the   phenomenology of  this model  is  quite different   from that of the  MSSM. 
In the Higgs sector, a sizable ($\sim10\%$) invisible branching ratio for Higgs decays into neutrinos and gravitinos is possible, together with small deviations in the Higgs couplings to gluons and photons, due to loop effects if the stop $\tilde{t}_R$ is light. 
These effects are not yet favored nor disfavored by the present LHC Higgs data, but could be seen in the near future 
by measuring a reduction of the visible Higgs BRs.
Higgsinos are absent in this model, and 
gauginos must  get Dirac masses   above the TeV.
Only  third-generation squarks  are required, by naturalness, to be below the TeV.
We have shown that the $R$-symmetry  implies that    squarks  decay mainly into quarks and either leptons or gravitinos. 
Therefore, evidence for models with the Higgs as a neutrino superpartner   can  be sought through the ongoing searches for events with third-generation quarks and missing energy (tailored for the MSSM with a massless neutralino) or through leptoquark searches for final states with heavy quarks and leptons.  In the stop decays into tops and neutrinos,
 the determination of the  top helicity  will be crucial to   unravel these scenarios. 
 
 Finally, if first and second generation squarks or sleptons are light enough, they can leave, via 3-body decays, interesting signatures at the LHC that deserve further study.

\vskip0.5cm
{\bf Note added:} While this work was being finalized, Ref.~\cite{Frugiuele:2012kp}
 appeared where   some of the squarks phenomenology of these models
is also  discussed.

%%%%%%%%%%%%%%%%%%%%%%%%%%%%%%%%%%%%%%%%%%%%%%%%%%%%%%%%%%%%%%%%%%%%%%%%%%%%%%%%%%%%%%%%%%%%%%%%%%%%%%%%%%%%%%%%%%%%%%%%%%%%%%%%%%%
\section*{Acknowledgements}
We are grateful to M.~Montull for help in relation to the fit of Fig.~\ref{Fig:bestfit}. FR acknowledges support from the Swiss National Science Foundation, under the Ambizione grant PZ00P2\_136932.
The work of AP was partly supported by the projects FPA2011-25948, 2009SGR894 and ICREA Academia Program.

%%%%%%%%%%%%%%%%%%%%%%%%%%%%%%%%%%%%%%%%%%%%%%%%%%%%%%%%%%%%%%%%%%%%%%%%%%%%%%%%%%%%%%%%%%%%%%%%%%%%%%%%%%%%%%%%%%%%%%%%%%%%%%%%%%%
\section*{Appendix:    \\
Possible  UV completions of   Higgsinoless models}

Here we want to briefly discuss  two  possible UV completion of  the model proposed in this article.
The first possibility   corresponds to the $R$-symmetric MSSM  of ref.~\cite{Frugiuele:2011mh}.
This model  contains two extra  Higgs superfields w.r.t. our model, $H_u$ and  $R_{d}$,  with a
  supersymmetric mass given by $\int d^2\theta\ \mu H_uR_d$.
  As in the MSSM the superfield $H_u$ can have Yukawa  terms with the up-quark sector,
$\int d^2\theta\ y_u H_uQU$, and   mix with $H$ (called $L_a$ in \cite{Frugiuele:2011mh})
  via a bilinear ($B_\mu$) soft-term, that we   write as  $\int d^4\theta\   R_d H^\dagger X^\dagger / M$.
For  $\mu\gg v$, we can integrate out $H_u$  and $R_d$, generating   the coupling \eq{topmass},
with  the identification $\mu=\Lambda$, or equivalently $Y_u=y_uF/(\mu  M)$.
Unfortunately, in this limit also a soft-term for $H$ is generated  at tree-level, $m_H\simeq F/ M$
that, for $Y_u\sim 1$, implies $y_u \sim \mu/m_H$.
Consequently,        $\mu>m_H$
leads to $y_u>1$,
thus possibly  leading to strong dynamics   slightly above the TeV. 
To extrapolate to higher energies we   could  assume,  along the lines of  \cite{Craig:2011ev}, that  
 the Higgs or top  are composite states of a strong group and use Seiberg dualities. 
The procedure of integrating out $H_u$ also generates corrections to the Higgs couplings that we  can explicitly calculate.
At  ${\cal O}(m^2_h/\mu^2)$,  we  find  that  only the Higgs coupling to the top is modified:
\begin{equation}
\frac{g_{htt}}{g_{htt}^{SM}}\simeq 1+2\frac{m_h^2}{\mu^2}\, .
\label{topmodi}
\end{equation}

Another possible UV completion of our model corresponds to a situation in which either the left-handed or
right-handed top  is  partly arising from  a  vector superfield.
For example, we can  have a massive vector superfield $V_\pm$, transforming under the SM as  a $(\bf 3,2)+(\bf \bar 3,2)$, and
with the following couplings: $\int d^4\theta [M_V^2V_+V_-+g_VV_-X^\dagger Q+g_VV_+H^\dagger U]$. Integrating out
$V_{\pm}$ gives   \eq{topmass}
with  $ y_u\sim g_V^2$ and $\Lambda\sim M\sim M_V$.
A soft-mass for $Q$  of order $m_Q\sim g_V F/M_V\sim Y_uM_V/g_V$ is also generated at tree-level 
and requires $g_V>1$  if we want $m_Q<M_V\sim $ TeV. 
Theories of massive gauge bosons, however,  need to be UV completed at energies $\sim 4\pi M_V/g_V$
either by incorporating them into a new  strong sector  or  by a Higgs mechanism.
In the second case,  the vector $V_{\pm}$ must be promoted into gauge bosons.
 A possibility   
discussed in \cite{Cai:2008ss} 
is to have $V_{\pm}$  arising from an  SU(5) gauge model. 
Notice that, as proposed in   \cite{Cai:2008ss}, we could take the limit in which the squarks are heavier than $M_V$
and have the vector  component of $V_+$  to be the main superpartner of the $t_L$ that would be in this case mainly a
gaugino.  Other options are given in \cite{Barbieri:1988im}.

\end{document}